\documentclass[preprint]{aastex}

\def\Msunh{~h^{-1}M_{\odot}\ }

\def\Mpch{h^{-1}{\rm Mpc}}
\def\kpch{h^{-1}{\rm kpc}}

\def\vc#1{\mathbf{#1}}
\def\lesssim{{_ <\atop{^\sim}}}

\begin{document}

\title{Numerical Simulations in Cosmology II: Spatial and Velocity Biases}

\author{Anatoly Klypin}
\affil{Department of Astronomy, New Mexico State University, Las
Cruces, NM 88001}

\begin{abstract}

We give a summary of recent results on spatial and velocity biases in
cosmological models. Progress in numerical techniques made it possible
to simulate halos in large volumes with a such accuracy that halos
survive in dense environments of groups and clusters of galaxies.
Halos in simulations look like real galaxies, and, thus, can be used to
study the biases -- differences between galaxies and the dark matter.
The biases depend on scale, redshift, and circular velocities of
selected halos. Two processes seem to define the evolution of the
spatial bias: (1) statistical bias and  (2) merger bias
(merging of galaxies, which happens preferentially in groups, reduces
the number of galaxies, but does not affect the clustering of the dark
matter). There are two kinds of velocity bias.  The pair-wise velocity
bias is $b_{12}=0.6-0.8$ at $r< 5h^{-1}$Mpc, $z=0$. This bias mostly
reflects the spatial bias and provides almost no information on the
relative velocities of the galaxies and the dark matter. One-point
velocity bias is a better measure of the velocities. Inside clusters
the galaxies should move slightly faster ($b_v =1.1-1.3$) than the dark
matter. Qualitatively this result can be understood using the Jeans
equations of the stellar dynamics. For the standard LCDM model we find
that the correlation function and the power spectrum of galaxy-size
halos at $z=0$ are antibiased on scales $r<5h^{-1}$Mpc and
$k\approx(0.15-30)h$Mpc$^{-1}$.

\end{abstract}

\section{Introduction}

The distribution of galaxies is likely biased with respect to the dark
matter. Therefore, the galaxies can be used to probe the matter
distribution only if we understand the bias. Although the problem of
bias has been studied extensively in the past (e.g., Kaiser 1984; Davis
et al., 1985; Dekel \& Silk 1986), new data on high redshift clustering
and the anticipation of coming measurements have recently generated
substantial theoretical progress in the field. The breakthrough in
analytical treatment of the bias was the paper by Mo \& White (1996),
who showed how bias can be predicted in the framework of the extended
Press-Schechter approximation. More elaborate analytical treatment has
been developed by Catelan et al. (1998ab), Porciani et al.(1998), and
Sheth \& Lemson (1998). Effects of nonlinearity and stochasticity were
considered in Dekel \& Lahav (1998) (see also \citet{TaruyaSuto}).

Valuable results are produced by ``hybrid'' numerical methods in which
low-resolution N-body simulations (typical resolution $\sim 20$kpc) are
combined with semi-analytical models of galaxy formation
\citep[e.g.][]{ Diaferio, Benson, Somerville}. Typically, results of
these studies are very close to those obtained with brute-force
approach of high-resolution ($\lesssim 2$kpc) N-body simulations
\citep[e.g.,][]{Colina, Ghignab}. This agreement is quite remarkable
because the methods are very different. It may indicate that the biases
of galaxy-size objects are controlled by the random nature of
clustering and merging of galaxies and by dynamical effects, which
cause the merging, because those are the only common effects in those
two approaches.

Direct N-body simulations can be used for studies of the biases
only if they have very high mass and force resolution. Because of
numerous numerical effects, halos in low-resolution simulations do not
survive in dense environments of clusters and groups (e.g., Moore, Katz 
\& Lake 1996; Tormen, Diaferio \& Syer, 1998; Klypin et al., 1999).
Estimates of the needed resolution are given in Klypin et
al. (1999). Indeed, recent simulations, which have sufficient resolution 
have found hundreds of  galaxy-size halos moving inside
clusters (Ghigna et al., 1998; Col\'in et
al., 1999a; Moore et al., 1999; Okamoto \& Habe, 1999).

It is very difficult to make accurate and trustful predictions of
luminosities for galaxies, which should be hosted by dark matter
halos. Instead of luminosities or virial masses we suggest to use
circular velocities $V_c$ for both numerical and observational
data. For a real galaxy its luminosity tightly correlate with the the
circular velocity. So, one has a good idea what is the circular
velocity of the galaxy. Nevertheless, direct measurements of circular
velocities of a large complete sample of galaxies are extremely
important because it will provide a direct way of comparing theory and
observations.
This lecture is mostly based on  results presented in Col\'in et
al. (1999ab) and  Kravtsov \& Klypin (1999).

\section{Oh, Bias, Bias}

There are numerous aspects and notions related with the bias. One
should be really careful to understand what what type of bias is
used. Results can be dramatically different. We start with introducing
the overdensity field. If $\bar\rho$ is the mean density of some
component (e.g., the dark matter or halos), then for each point
$\vc{x}$ in space we have  $\delta(\vc{x})\equiv
[\rho(\vc{x})-\bar\rho]/\bar\rho$. The overdensity can be decomposed
into the Fourier spectrum, for which we can find the power spectrum
$P(k)=\langle |\delta_{\vc{k}}|^2\rangle$. We then can find the
correlation function $\xi(r)$ and the rms fluctuation of $\delta(R)$
smoothed on a given scale $R$. We can construct the statistics for each
component: dark matter, galaxies, or halos with given properties. Each
statistics gives its own definition of bias $b$:

\begin{equation}
P_h(k) =b_P^2 P_h(k) , \qquad \xi_h(r)= b_{\xi}^2 \xi_{\rm dm}(r), \qquad 
\delta_h(R) =b_{\delta}\delta_{\rm dm}(R).
\end{equation}

The three estimates of the bias $b$ are related. In special case, when
the bias is
linear, local, and scale independent  all three forms of bias are all equal. In general case
they are different and they are complicated nonlinear functions of scale, mass of
the halos or galaxies, and redshift. The dependence on the scale is not
local in the sense that the bias in a given position in space may depend
on environment (e.g., density and velocity dispersion) on a larger
scale. Bias has memory: it depends on the local history of fluctuations.
There is another complication: bias very likely is not a deterministic
function. One source of this stochasticity is that it is
nonlocal. Dependence on the history of clustering may also introduce
some random effect.

There are some processes, which we know create and affect the bias. At
high redshifts there is statistical bias: in a Gaussian correlated
field high density regions are more clustered than the field itself
\citep{Kaiser}.  \citet{MoWhite} showed how the
extended Press-Schechter formalism can be used for derivation of the
bias of the dark matter halos. In the limit of small perturbations on
large scales the bias is \citep{Catelana, TaruyaSuto}

\begin{equation}
b(M,z,z_f) = 1 + \frac{\nu^2 -1}{\delta_c(z,z_f)}.
\end{equation}

\noindent
Here $\nu=\delta_c(z,z_f)/\sigma(M,z)$ is the relative amplitude of a
fluctuation on scale $M$ in units of the rms fluctuation
$\sigma(M,z)$ of the density field at redshift $z$. Parameter $z_f$ is
the redshift of halo formation. The critical threshold of the top-hat
model is $\delta_c(z,z_f)=\delta_{c,0}D(z)/D(z_f)$, where $D$ is the
growth factor of perturbations and $\delta_{c,0}=1.69$. At high
redshifts, parameter $\nu$ for galaxy-size fluctuations is very large
and $\delta_c$ is small. As the result, galaxy-size halos are expected
to be more clustered (strongly biases) as compared to the dark
matter. The bias is larger for more massive objects. As fluctuations
grow, new forming galaxy-size halos do not represent as high peaks as
at large redshifts and the bias tends to decrease. It also looses its
sensitivity. 

At later stages another process start to change the bias. In groups and
cluster progenitors the merging and destruction of halos reduces the
number of halos. This does not happen in the field where number of
halos of given mass may only increase with time. As the result, the
number of halos inside groups and cluster progenitors is reduced
relatively to the field. This produces (anti)bias: there is
relatively smaller number of halos as compared with the dark
matter. This merging bias does not depend on mass of halos and it has a
tendency to slow down once a group becomes a cluster with large
relative velocity of halos \citep{KravtsovKlypin}. 
 
Here is a list of different types of biases. We classify them into
three groups: (1)  measures of bias 
(2) terms related with the description of biases, (3)
physical precesses, which produce or change the bias.

\begin{itemize}
\item Measures of bias
     \begin{enumerate}
		\item bias measured in a statistical sense (e.g., ratio of
                         correlation functions $\xi_h(r) =b^2\xi_{\rm dm}(r)$)
            \item bias measured point-by-point (e.g.,
				$\delta_h(\vc{x})-\delta_m(\vc{x})$ diagrams) 
     \end{enumerate}
\item Description of biases
     \begin{enumerate}
		\item local and nonlocal bias. For example,
             $b(R)=\sigma_h(R)/\sigma_m(R)$ is the local bias. If $b=b(R;\tilde R)$,
             the bias is nonlocal, where $\tilde R$ is some other scale or scales.
		\item linear and nonlinear bias. If in $\xi_h(r) =b^2\xi_{\rm
			dm}(r)$ the bias $b$ does not depend on $\xi_{\rm dm}$, it is
                  the linear bias. 
		\item scale dependent and scale independent bias. If $b$ does not
depend on scale at which the bias is estimated, the bias is scale
independent. Note that in general, the bias can be nonlinear and scale
independent, but this highely nonlikely.
	      \item stochastic and deterministic.
     \end{enumerate}
\item Physical precesses, which produce or change the bias
	\begin{enumerate}
		\item statistical bias. Bias, which arises when a specific subset
                 of points is selected from a Gaussian field.
		\item merging bias. Bias produced due to merging and destruction
				of halos.
 		\item physical bias. Any bias due to physical processes inside
                          forming galaxies.
     \end{enumerate}
\end{itemize}

\section{Spatial bias}

\citet{Colina} have simulated different cosmological models and using
the simulations studied halo biases. Most of the results presented here
are for  currently favored  $\Lambda$CDM model with the
following parameters:
$\Omega_0=1-\Omega_{\Lambda}=0.3$, $h=0.7$, $\Omega_b=0.032$,
$\sigma_8=1$. The model was simulated with $256^3$ particles in
a 60$h^{-1}$Mpc box. Formal mass and force resolutions are
$m_1=1.1\times 10^9h^{-1}M_{\odot}$ and $2h^{-1}$kpc. Bound Density
Maximum halo finder was used to identify halos with at least 30 bound
particles. For each halo we find maximum circular velocity
$V_c=\sqrt{GM(<r)/r}$.

\begin{figure}[tb!] 
\plotone{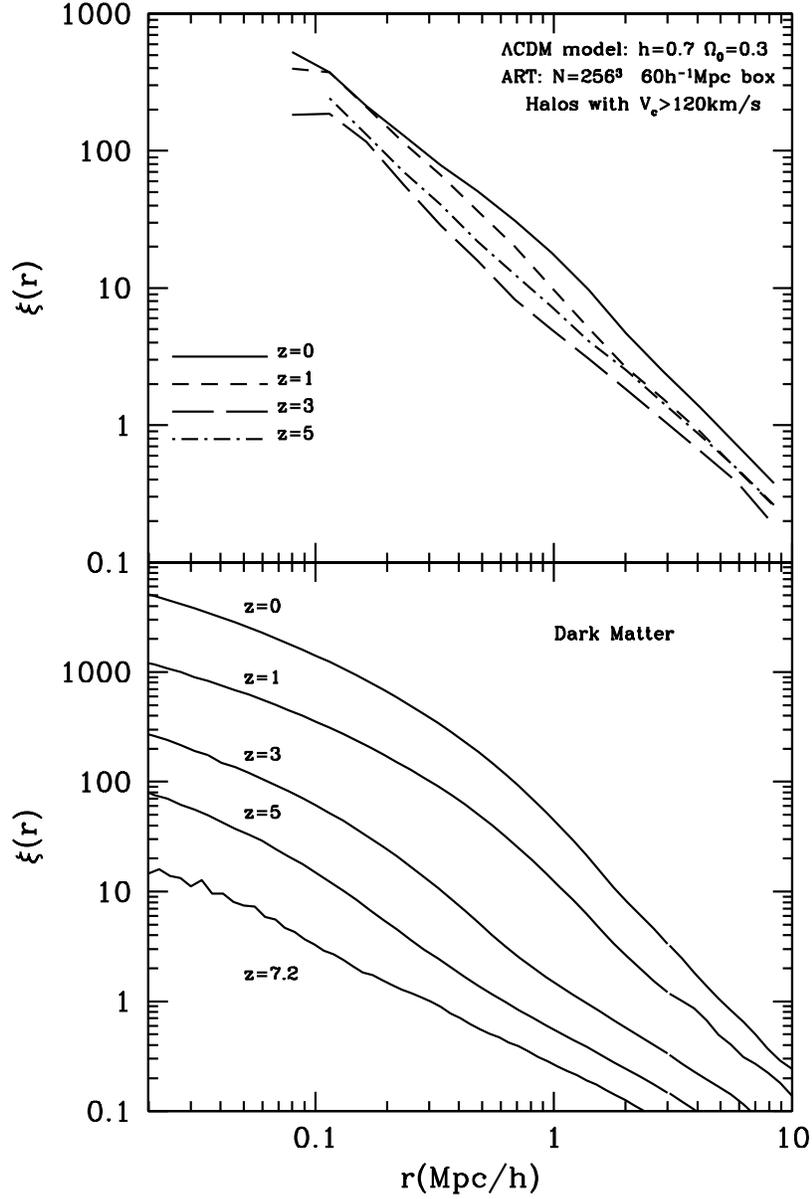}
\caption{\small Evolution of the correlation function of the dark matter and
halos. Correlation function of the dark matter increases monotonically
with time. At any given moment it is not a power law. The correlation
function of halos is a power-law, but it is not monotonic in time} 
\label{fig-CE}
\end{figure}

\begin{figure}[tb!] 
\epsscale{1.05} 
\plottwo{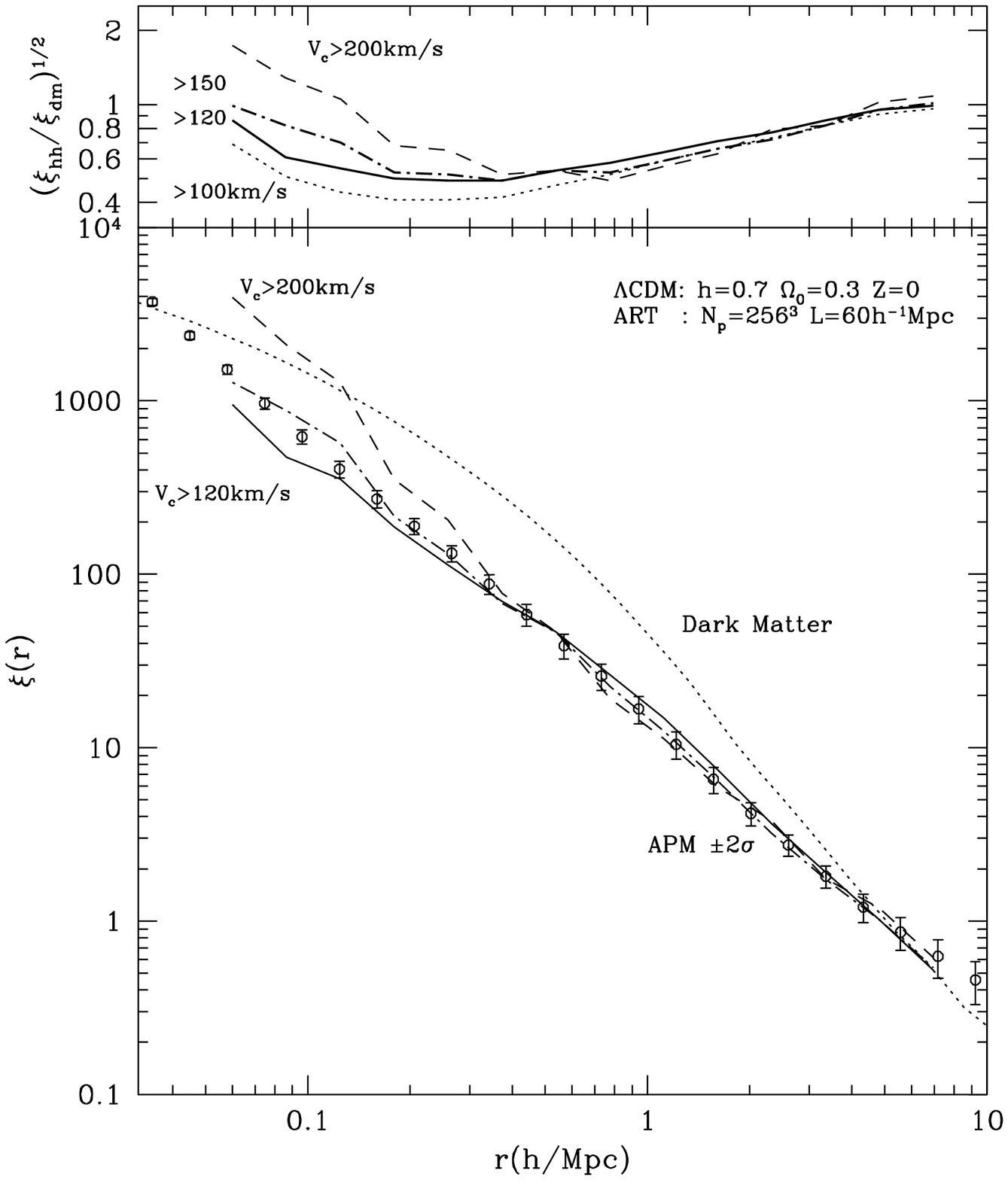}{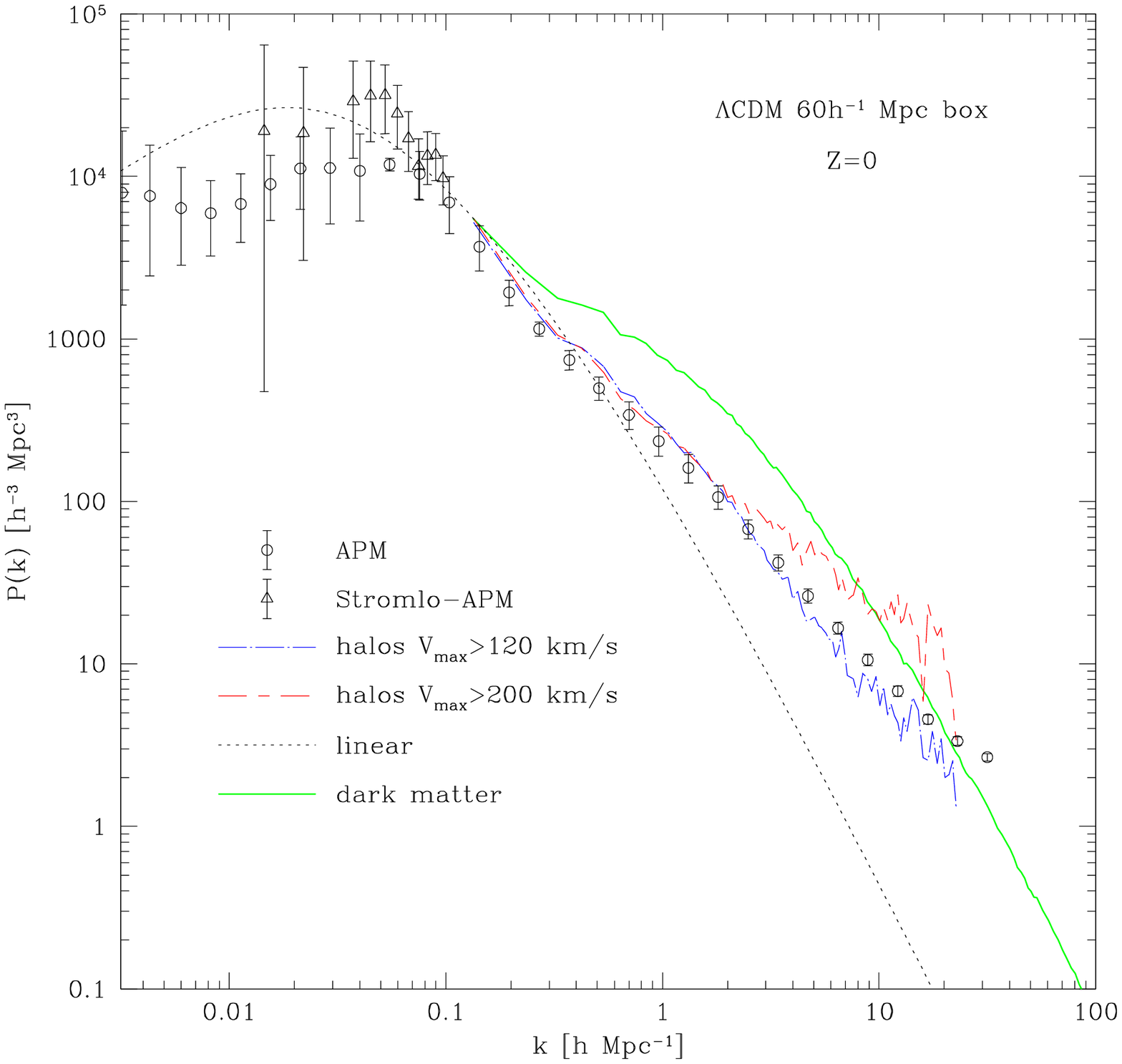}
\caption{\small The correlation function and the power spectrum of
halos of different limiting circular velocities in the $\Lambda$CDM
model. Results are compared with the observational data from the APM
and Stromlo-APM surveys. Bias is scale dependent, but it does  not depend
much on halo mass} \label{fig-1}
\end{figure}

In figure~\ref{fig-CE} we compare the evolution of the correlation
functions of the dark matter and halos. There are  remarkable
differences between halos and the dark matter. The correlation
functions of the dark matter always increases with time (but the rate
is different on different scales) and it never is a power-law. The correlation
functions of the halos at redshifts goes down and then starts to
increase again. It is accurately described by a power-law with slope
$\gamma=(1.5-1.7)$. 
Figure~\ref{fig-1} presents a comparison of the theoretical and
observational data on correlation functions and power spectra. The dark
matter clearly predicts much too high amplitude of clustering. The
halos are much closer to the observational points and predict
antibias. For the correlation function the antibias appears on scales $r
<5h^{-1}$Mpc; for the power spectrum the scales are $k>0.2h{\rm
Mpc}^{-1}$. One may get an impression that the antibias starts at
longer waves in the power spectrum $\lambda =2\pi/k\approx 30h^{-1}{\rm
Mpc}$ as compared with $r\approx 5h^{-1}$Mpc in the correlation
function. There is no contradiction: sharp bias at small distances in the
correlation function when Fourier transformed to the power spectrum
produces antibias at very small wavenumbers. Thus, the bias should be
taken into account at long waves when dealing with the power spectra.
There is an inflection point in the power spectrum where the nonlinear
power spectrum start to go upward (if one moves from low to high
$k$) as compared with the prediction of the linear theory. Exact
position of this point may have been affected by the finite size of the 
simulation box $k_{\rm min}=0.105h^{-1}$Mpc, but effect is expected to
be small.

\begin{figure}[tb!] 
\epsscale{0.8} 
\plotone{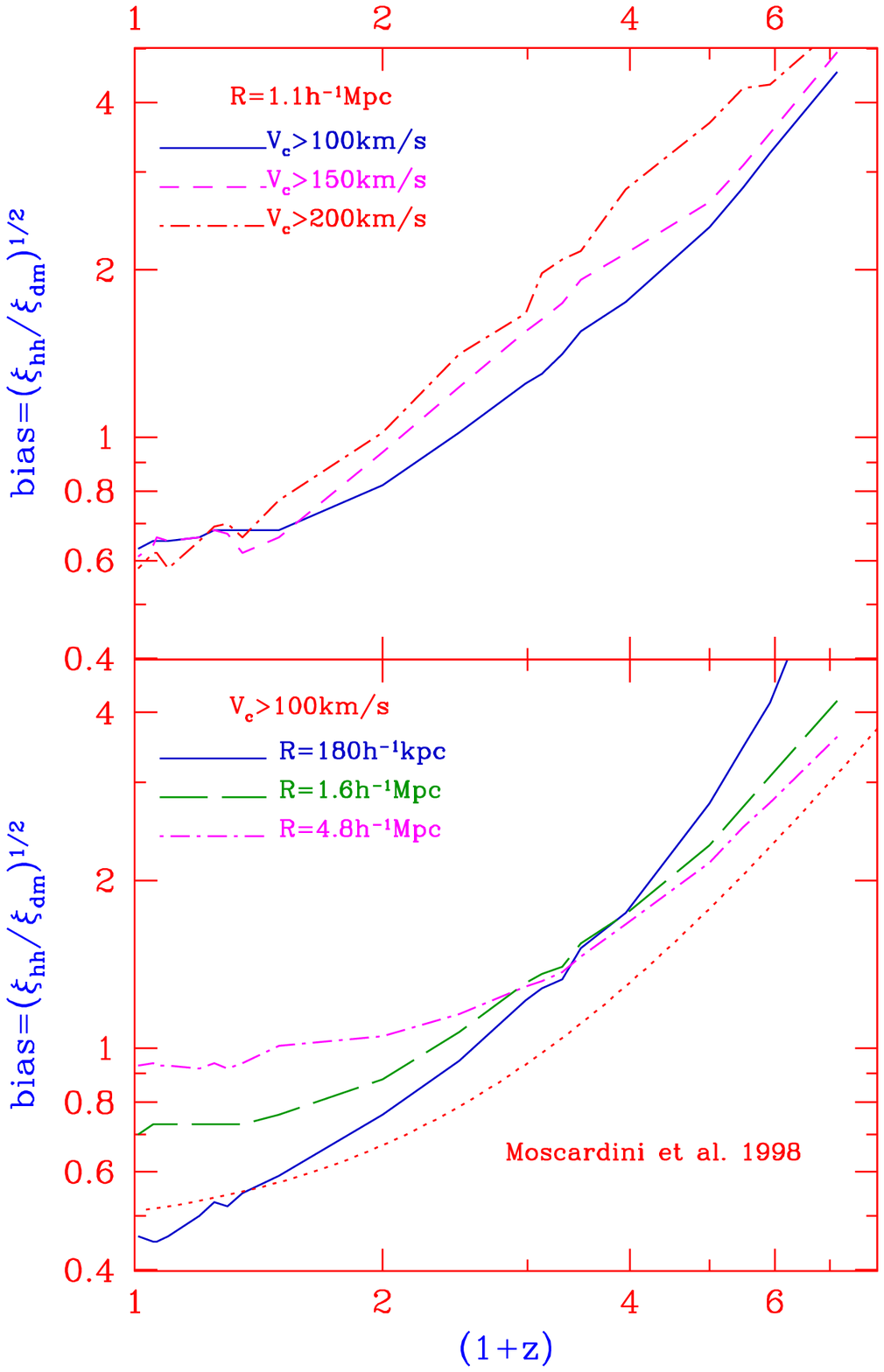}
\caption{\small {\it Top panel:} The evolution of bias at comoving
scale of $0.54\Mpch$ for halos with different circular velocity.  
{\it Bottom panel:} Dependence of the  bias on the scale for halos with
the same circular velocity.} \label{fig-bias}
\end{figure}

\begin{figure}[tb!] 
\epsscale{0.8} 
\plotone{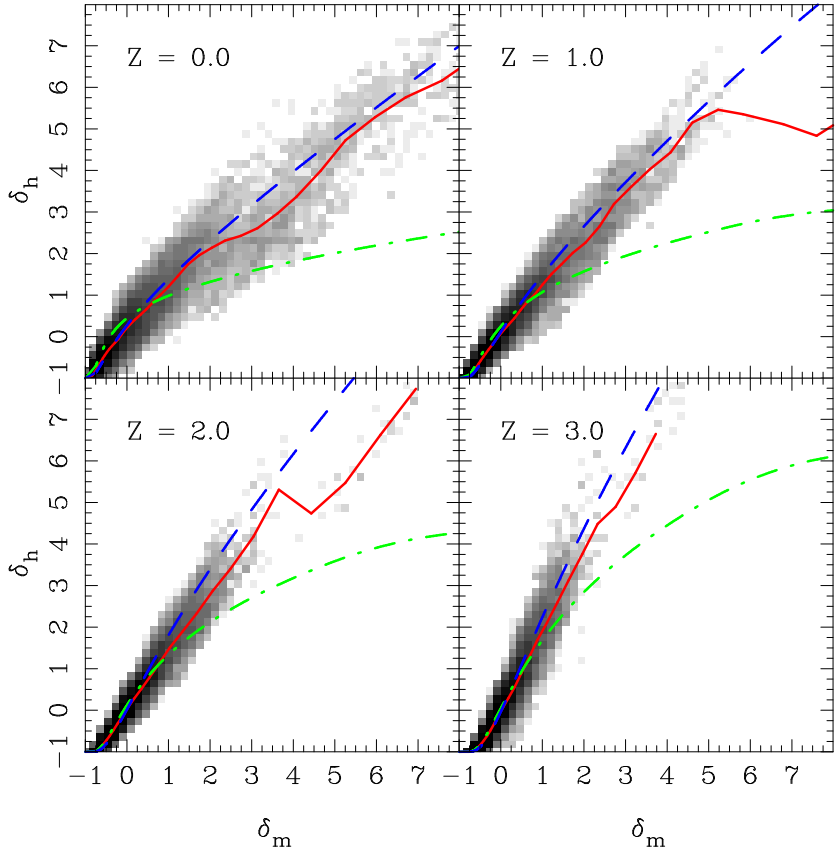}
\caption{\small Overdensity of halos $\delta_h$ vs. the overdensity of the
dark matter $\delta_m$. The overdensities are estimated in spheres of
radius $R_{\rm TopHat}=5h^{-1}$Mpc. Intensity of the grey shade
corresponds to the natural logarithm of the number of spheres in a 2D
grid in $\delta_h$- $\delta_m$ space. The solid curves show the average
relation. The dot-dashed curve is a prediction of an analytical model,
which assumes that formation redshift $z_f$ of halos coincides with
observation redshift (typical assumption for the Press-Schechter
approximation). The long-dashed curve is for model, which assumes that
substructure survives for some time after it falls into a larger object: 
$z_f=z+1$} \label{fig-2}
\end{figure}

\begin{figure}[tb!] 
\epsscale{0.8} 
\plotone{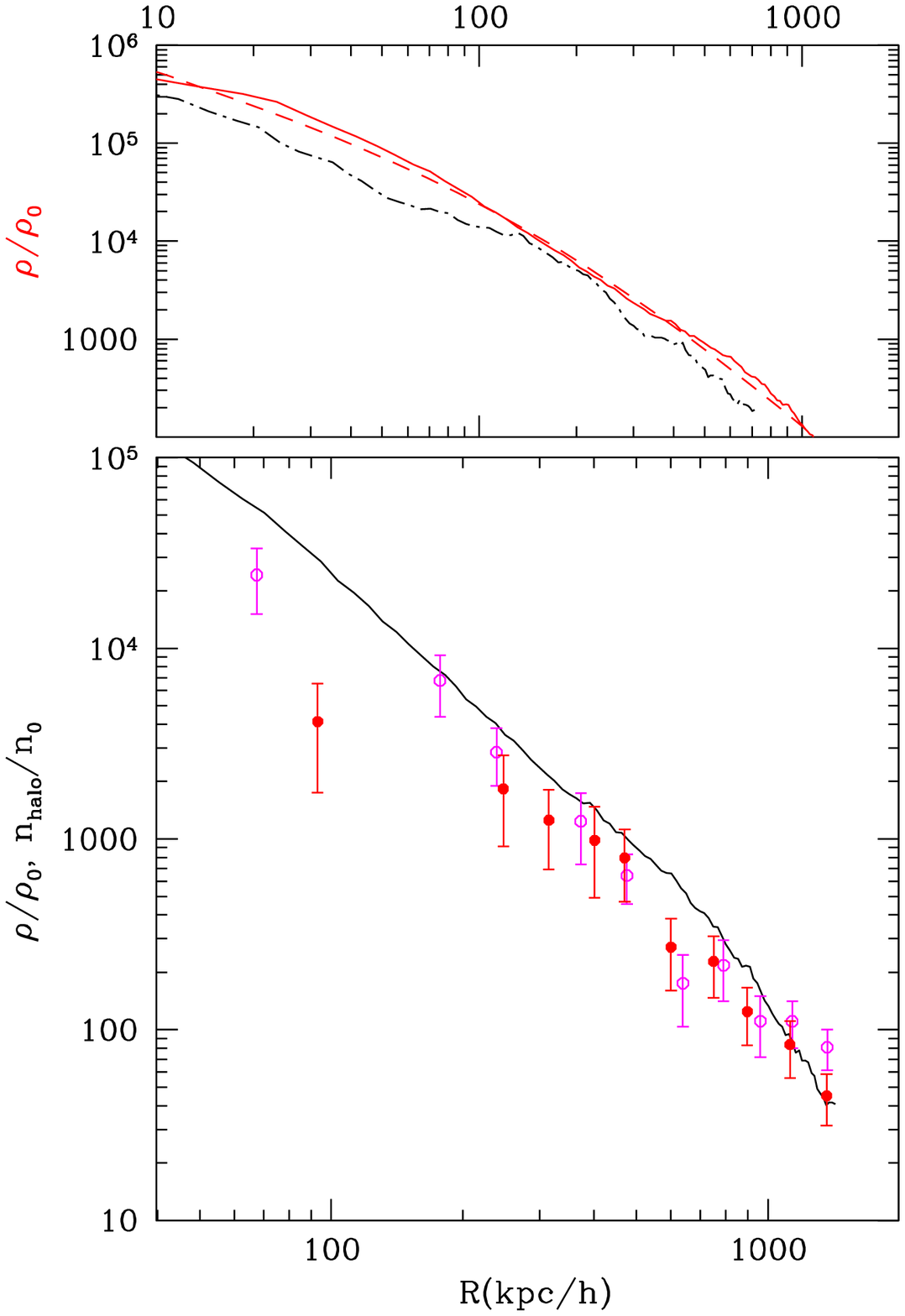}
\caption{\small Density profiles for a cluster with mass $2.5\times
10^{14}\Msunh$. {\it Top panel:} Dark matter density in units of the
mean matter density at $z=0$ (solid line) and at $z=1$ (dot-dashed
line). The Navarro-Frenk-White profile (dashed line)  provides a very
good fit at $z=0$.  The $z=1$ profile is given in proper (not comoving)
 units. {\it Bottom panel:} Number density profiles of halos in the
cluster at  $z=0$. (solid circles) and at $z=1$ (open circles) as
compared with the $z=0$ dark matter profile (solid curve). There is
antibias on scales below $300\kpch$.}
\label{fig-cluster}
\end{figure}

At $z=0$ the bias almost does not depend on the mass limit of the
halos. There is a tendency of more massive halos to be more clustered
at very small distances $r<200h^{-1}$kpc, but at this stage it is not
clear that this is not due to residual numerical effects around centers
of clusters. The situation is different at high redshift. At very high
redshifts $z>3$ galaxy-size halos are very strongly (positively)
biased. For example, at $z=5$ the correlation function of halos with
$v_c>150{\rm km}/{\rm s}$ was 15 times larger than that of the dark
matter at $r=0.5h^{-1}$Mpc (see Fig.8 in \citet{Colina}). The
bias was also very strongly mass-dependent with more massive halos
being more clustered. At smaller redshifts the bias was quickly
declining. Around $z=1-2$ (exact value depends on halo circular
velocity) the bias crossed unity and became less than unity (antibias)
at later redshifts.

Evolution of bias is illustrated by Figure~\ref{fig-2}. The figure
shows that at all epochs the overdensity of halos tightly correlates with
the overdensity of the dark matter. The slope of the relation depends on
the dark matter density and evolves with time. At $z>1$ halos are
biased ($\delta_h > \delta_m$) in overdense regions with $\delta_m>1$ and
antibiased in underdense regions with $\delta_m< -0.5$ At low redshifts
there is an antibias at large overdensities and almost no bias at low
densities.

Figure~\ref{fig-cluster} shows the density profiles for a cluster with mass $2.5\times
10^{14}\Msunh$. There is
antibias on scales below $300\kpch$. This is an example of the merging
and destruction bias. Some of the halos have merged or were destroyed
by the central cD halo of the cluster. As the result, there is smaller
number of halos in the central part as compared with what we would
expect if the number density of halos follower the density of the dark
matter (the full curve in the bottom panel). Note that in the outer
parts of the cluster halos closely follow the dark matter.

\section{Velocity bias}
\begin{figure} [tb!] 
\epsscale{1.05}
\plottwo{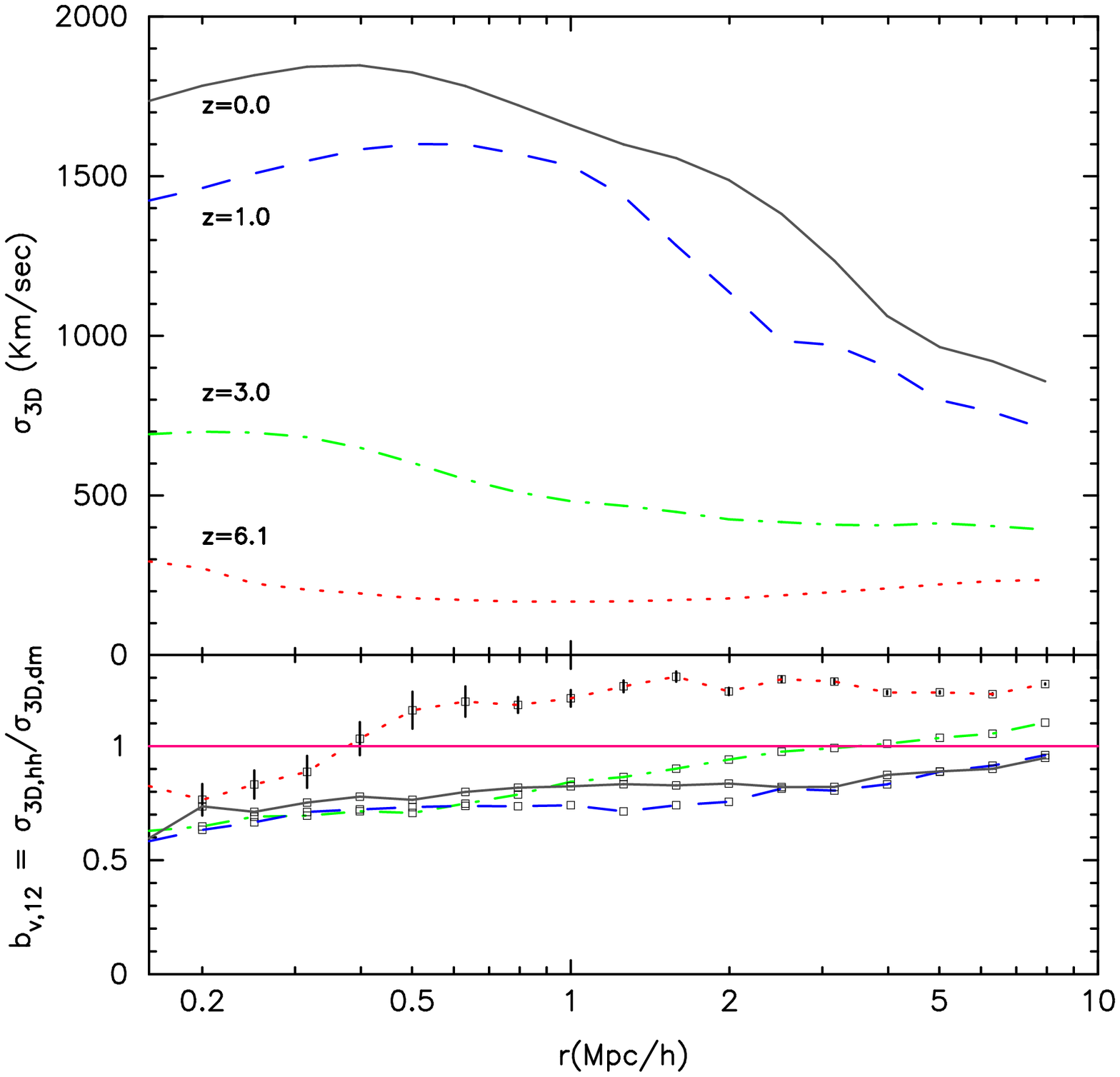}{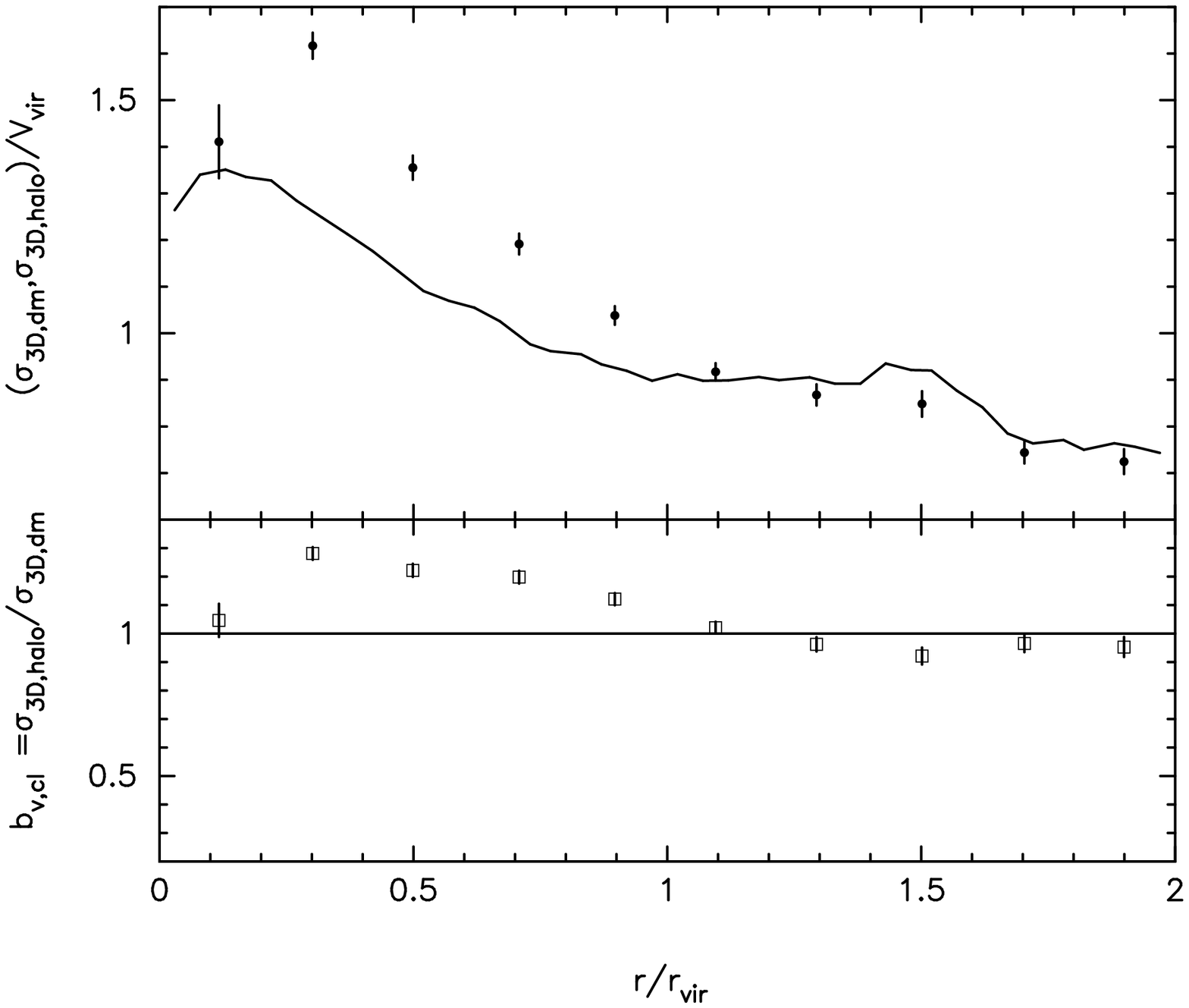}
\caption{\small {\it Left:} Two-point velocity bias. {\it Right:}  {\it Top panel:} 3D
 rms velocity for halos (circles) and for dark matter (full curve) in
 twelve largest clusters.  {\it Bottom panel:} velocity bias in the
 clusters. The bias in the first point increases to 1.2 if the central
 cD halos are excluded from analysis. Errors correspond to 1-sigma
 errors of the mean obtained by averaging over 12 clusters at two
moments of time. Fluctuations for individual clusters are larger.}
\label{fig-3}
\end{figure}

There are two statistics, which measure velocity biases -- differences
in velocities of the galaxies (halos) and the dark matter. For a review
of results and references see Col\'in et al. (1999). Two-particle or
pairwise velocity bias (PVB) measures the relative velocity dispersion
in pairs of objects with given separation $r$: $b_{12}=\sigma_{\rm
halo-halo}(r)/\sigma_{\rm dm-dm}(r)$. Figure~\ref{fig-3} (left panel)
shows this bias.  It is very sensitive to the number of pairs inside
clusters of galaxies, where relative velocities are largest. Removal of
few pairs can substantially change the value of the bias. This
``removal'' happens when halos merge or are destroyed by central
cluster halos.

 One-point velocity bias is estimated as a ratio
of the rms velocity of halos to that of the dark matter:
$b_1=\sigma_{halos}/\sigma_{dm}$. It is typically applied to clusters
of galaxies where it is measured at different distances from the
cluster center. For analysis of the velocity bias in clusters \citet{Colinb}
have selected twelve most massive clusters in a simulation of the
$\Lambda$CDM model.  The most massive cluster had virial mass $6.5\times
10^{14}h^{-1}M_{\odot}$ comparable to that of the Coma cluster. The
cluster had 246 halos with circular velocities larger than
90~km/s. There were three Virgo-type clusters with virial masses in the
range $(1.6-2.4)\times 10^{14}h^{-1}M_{\odot}$ and with approximately
100 halos in each cluster.  Just as the spatial bias, PVB is positive
at large redshifts (except for the very small scales) and decreases
with the redshift. At lower redshifts it does not evolve much and stays
below unity (antibias) at scales below $5h^{-1}$Mpc on the level
$b_{12}\approx (0.6-0.8)$.

 Figure~\ref{fig-3} shows one-point velocity bias
in clusters at $z=0$. Note that the sign of the bias is now different:
halos move slightly faster than the dark matter. The bias is stronger
in the central parts $b_1=1.2-1.3$ and goes to almost no bias
$b_1\approx 1$ at the virial radius and above. Both the antibias in the
pairwise velocities and positive one-point bias are produced by the
same physical process -- merging and destruction of halos in central
parts of groups and clusters. The difference is in the different
weighting of halos in those two statistics. Smaller number of
high-velocity pairs significantly changes PVB, but it does not affect much 
the one-point bias because it is normalized to the number of halos at
a given distance from the cluster center. At the same time, merging
preferentially happens for halos, which move with smaller velocity at a 
given distance from the cluster center. Slower halos have shorter
dynamical times and have smaller apocenters. Thus, they have better
chance to be destroyed and merged with the central cD halo. Because
low-velocity halos are eaten up by the central cD, velocity dispersion
of those, which survive, is larger. Another way of addressing the issue 
of the velocity bias is to use the Jeans equations. If we have a tracer 
population, which is in equilibrium in a potential produced by mass
$M(<r)$, then
\begin{equation}
-r\sigma_r^2(r)\left[\frac{d\ln\sigma_r^2(r)}{d\ln r}+
\frac{d\ln\rho(r)}{d\ln r}+2\beta(r)\right] =GM(<r),
\end{equation}
\noindent where $\rho$ is the number density of the tracer, $\beta$ is
the velocity anisotropy, and $\sigma_r$ is the rms radial velocity. The
r.h.s. of the equation is the same for the dark matter and the
halos. If the term in the brackets would be the same, there would be no
velocity bias. But there is systematic difference between the halos
and the dark matter: the slope of the distribution halos in a cluster
$\frac{d\ln\rho(r)}{d\ln r}$ is smaller than that of the dark matter (see
Col\'in et al., 1998, Ghigna et al., 1999). The reason for the
difference of the slopes is the same --
merging with the central cD. Other terms in the equation also have
small differences, but the main contribution comes from the slope of
the density. Thus, as long as we have spatial antibias of the halos,
there should be a small positive one-point velocity bias in clusters
and a very strong antibias in pairwise velocity. Exact values of the
biases are still under debate, but one thing seems to be certain: one
bias does not go without the other.

\begin{figure}[tb!] 
\epsscale{0.5}
\plotone{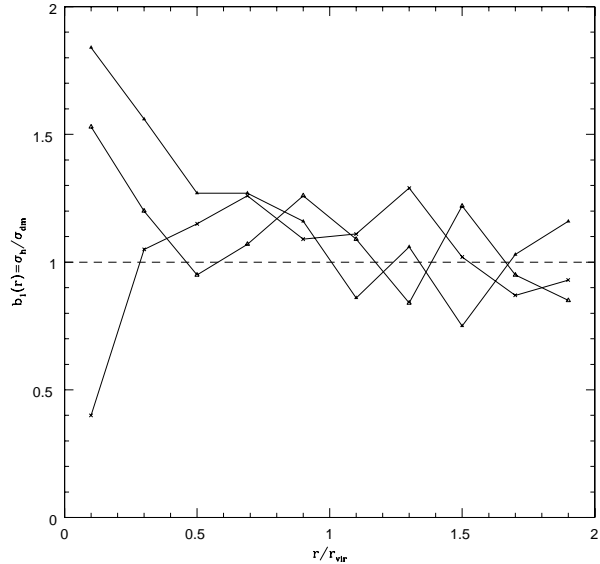}
\caption{\small 
One-point velocity bias for 3 Virgo-type clusters in the
simulation. Central cD halos are not included. Fluctuations in the bias
are very large because each cluster has only $\sim$100 halos with $V_c>
90km/s$ and because of substantial substructure in the clusters. }
\label{fig-b12}
\end{figure}

The velocity bias in clusters is difficult to measure because it is
small. The Figure~\ref{fig-3} may be misleading because
it shows the average trend, but it does not give the level of
fluctuations for a single cluster. Note that the errors in the plots
correspond to the error of the mean obtained by averaging over 12
clusters and two close moments of time. Fluctuations for a single
cluster are much larger. Figure~\ref{fig-3} shows results for three
Virgo-type clusters in the simulation. The noise is very large because
of both poor statistics (small number of halos) and the noise produced
by residual non-equilibrium effects (substructure). Comparable (but
slightly smaller) value of $b_v$ was recently found in simulations by
Ghigna et al. (1999, astro-ph/9910166) for a cluster in the same mass
range as in Figure~\ref{fig-3}. Unfortunately, it is difficult to make
detailed comparison with their results because Ghigna et al. (1999) use
only one hand-picked cluster for a different cosmological model. Very
likely their results are dominated by the noise due to residual
substructure. Results of another high-resolution simulation by Okamoto
\& Habe (1999) are consistent with our results.

\section{Conclusions}

There is a number of physical processes, which can contribute to the
biases. In our papers we explore dynamical effects in the dark matter
itself, which result in differences of the spatial and velocity
distribution of the halos and the dark matter. Other effects related to 
the formation of luminous parts of galaxies also can produce or change
biases. At this stage it is not clear how strong are those
biases. Because there is a tight correlation between the luminosity and
circular velocity of galaxies, any additional biases are limited by the 
fact that galaxies ``know'' how much dark matter they have. 

Biases in the halos are reasonably well understood and can be
approximated on a few Megaparsec scales by analytical models. 
We find that the biases in the distribution of the halos are sufficient 
to explain within the framework of standard cosmological models the
clustering properties of galaxies on a vast ranges of scales from
100~kpc to dozens Megaparsecs. Thus, there is neither need nor much
room for additional biases in the standard cosmological model. 

In any case, biases in the halos should be treated as benchmarks for
more complicated models, which include non-gravitational physics. If a
model can not reproduce biases of halos or it does not have enough
halos, it should be rejected, because it fails to have correct dynamics 
of the main component of the Universe -- the dark matter.



\end{document}